\documentclass[sigconf]{acmart}

\usepackage{tikz}
\usetikzlibrary{shapes.geometric, arrows.meta, positioning, fit, backgrounds}

\usepackage[table]{xcolor}
\usepackage{array}
\usepackage{colortbl}
\usepackage{makecell}
\usepackage{multirow}

\definecolor{TableHeader}{HTML}{EAF2FB}
\definecolor{TableStripe}{HTML}{F7FAFC}
\definecolor{TableAccent}{HTML}{DCEAF7}
\definecolor{TableSummary}{HTML}{EEF6EE}
\definecolor{TableEmph}{HTML}{FBEFEA}

\definecolor{TraceHeader}{HTML}{E8EEF7}
\definecolor{TraceSuccess}{HTML}{EAF6EE}
\definecolor{TraceWarn}{HTML}{FFF4E5}
\definecolor{TraceFail}{HTML}{FBEAEA}
\definecolor{TraceTotal}{HTML}{E8EAF6}

\definecolor{PsiLow}{HTML}{EAF6EE}
\definecolor{PsiMid}{HTML}{FFF4E5}
\definecolor{PsiHigh}{HTML}{FBEAEA}
\definecolor{PsiNeutral}{HTML}{F7FAFC}

\definecolor{ReliabilityHeader}{HTML}{EAF4F4}
\definecolor{ReliabilityStrong}{HTML}{EAF6EE}
\definecolor{ReliabilityCaution}{HTML}{FFF4E5}

\definecolor{RobustHeader}{HTML}{EFEAF7}
\definecolor{RobustStrong}{HTML}{EEEAF8}
\definecolor{RobustMedium}{HTML}{FFF4E5}

\definecolor{RedesignBefore}{HTML}{FBEAEA}
\definecolor{RedesignAfter}{HTML}{EAF6EE}
\definecolor{RedesignBest}{HTML}{E6F4EA}

\newcommand{\bestcell}[1]{\cellcolor{PsiLow}\textbf{#1}}
\newcommand{\midcell}[1]{\cellcolor{PsiMid}#1}
\newcommand{\highcell}[1]{\cellcolor{PsiHigh}\textbf{#1}}

\newcommand{\strongcell}[1]{\cellcolor{ReliabilityStrong}\textbf{#1}}

\newcommand{\included}{\cellcolor{TraceSuccess}\textbf{Included}}
\newcommand{\excluded}{\cellcolor{TraceFail}Excluded}
\newcommand{\flagged}{\cellcolor{TraceWarn}Flagged}

\renewcommand{\arraystretch}{1.12}

\usepackage{pgfplots}
\pgfplotsset{compat=1.18}
\usetikzlibrary{matrix,calc}
\AtBeginDocument{%
  }

\setcopyright{acmlicensed}
\copyrightyear{2018}
\acmYear{2018}
\acmDOI{XXXXXXX.XXXXXXX}
\acmConference[Conference acronym 'XX]{Make sure to enter the correct
  conference title from your rights confirmation email}{June 03--05,
  2018}{Woodstock, NY}
\acmISBN{978-1-4503-XXXX-X/2018/06}

\begin{document}

\title{The Privacy Placebo: Diagnosing Consent Burden through Performative Scrolling}

\author{Haoze Guo}
\email{hguo246@wisc.edu}
\orcid{0009-0009-5987-1832}
\affiliation{%
  \institution{University of Wisconsin - Madison}
  \city{Madison}
  \state{WI}
  \country{USA}
}

\author{Ziqi Wei}
\email{zwei232@wisc.edu}
\orcid{0009-0009-7541-8376}
\affiliation{%
  \institution{University of Wisconsin - Madison}
  \city{Madison}
  \state{WI}
  \country{USA}
}

\renewcommand{\shortauthors}{Guo et al.}

\begin{abstract}
While consent banners and privacy policies invite users to read and choose, many choices are shaped by interface structures that make non-accepting alternatives harder to discover or act upon. This paper uses \emph{performative scrolling} as a design-side lens for identifying slow, low-yield consent interactions in which meaningful alternatives may remain structurally delayed or obscured. We present the \emph{Performative Scrolling Index} (PSI), a reproducible interface-audit metric for measuring pre-choice burden before a meaningful non-accepting alternative becomes visible and actionable. PSI decomposes burden into four observable components: distance, time, focus loops, and hidden reveals. We treat the component vector as the primary audit output and use scalar PSI as a compact summary rather than as a psychological scale. We provide a least-effort audit protocol, implementation details, design-side invariants, worked trace examples, and a live deployment across 500 websites under desktop/mobile and pointer/keyboard traversal conditions. We further compare PSI against simpler baseline heuristics and test robustness across plausible weighting profiles. Together, these analyses show how structural choices such as offscreen alternatives, fragmented disclosure, staged modal flows, and keyboard fragility can increase pre-choice friction in ways that simple checklist metrics may miss. PSI is not a measure of comprehension or legal sufficiency; rather, it is a diagnostic of interface-side burden intended to support reproducible audits and redesigns.
\end{abstract}

\ccsdesc[500]{Security and privacy~Privacy protections}
\ccsdesc[300]{Security and privacy~Usability in security and privacy}
\ccsdesc[100]{Human-centered computing~HCI design and evaluation methods}
\ccsdesc[100]{Human-centered computing~Web-based interaction}

\keywords{consent management platforms, privacy UX, interaction logging, consent burden, friction metrics}

\maketitle

\section{Introduction}
Even with mounting evidence that people tend to not read privacy policies and generally accept them out of habit or resignation \citep{acquisti2015science,mcdonald2008cost}, notice-and-consent remains the dominant mechanism through which platforms seek to establish user consent to processing. In practice, however, consent becomes a series of rituals: a modal pops up, people scroll through a long passage of text/buttons that occupy their interaction, and there is ultimately an acceptance. The end result is an interaction that can signify diligence without necessarily changing what the user understands, acts on, or ultimately chooses.

This paper develops a design-side lens on that routine. We focus on \emph{performative scrolling}—a familiar, careful movement through text that resembles reading yet often functions as a coping ritual. The behavior is not irrational: declining can be effortful; legalistic language is hard to parse; the benefit of returning to one’s task typically outweighs the marginal benefit of parsing boilerplate; and the experience repeats across sites. Scrolling carefully allows people to reconcile two beliefs: “I ought to review my privacy choices,” and “I will likely accept anyway.” This motivates what we refer to as the \emph{privacy placebo}: a design-side framing for consent interactions that can appear to support careful choice while leaving meaningful alternatives structurally displaced or delayed \citep{barth2017privacyparadox}.

The primary difficulty is quantifying the disassociation between obtaining consent and exercising agency. Text length, number of scrolls, or time-on-task do not suffice. We therefore define a \emph{Performative Scrolling Index} (PSI), computable from minimally invasive telemetry on the \emph{interfaces themselves}: a scripted least-effort protocol goes through the flow; we log transition states and summarize the burden required \emph{before} a meaningful alternative becomes actionable. PSI does not directly measure comprehension or satisfaction; rather, it operationalizes pre-choice interaction friction that can plausibly support ritualized compliance (e.g., displaced alternatives, layered disclosure, and keyboard fragility).

This paper makes three contributions. First, we introduce \emph{performative scrolling} as a structural lens for understanding how consent interfaces can invite user actions that resemble careful review while contributing little to meaningful choice. Second, we present the \emph{Performative Scrolling Index} (PSI), a reproducible interface-audit metric that summarizes pre-choice burden before a meaningful alternative becomes actionable. PSI is computed from observable interface signals---distance, time, focus loops, and hidden reveals---rather than from inferred user intent or self-reported understanding. Third, we provide an audit protocol, implementation details, sensitivity analysis, a medium-scale live deployment, and an initial inter-rater reproducibility check demonstrating how PSI can be applied and inspected across device and traversal conditions.

We scope PSI deliberately. PSI is \emph{not} a measure of legal sufficiency, comprehension, satisfaction, or welfare. Instead, it is a design-side measure of \emph{pre-choice interaction friction}: the burden imposed before a non-accepting alternative is visible and actionable. This framing allows PSI to support reproducible audits and redesign discussions while leaving downstream questions of comprehension and legal validity to complementary methods.

\section{Related Work}
Interfaces can influence decision-making in ways that manipulate cognitive limitations, habituation, and uneven effort. Early work on malicious interface design and dark patterns established that interface structures can systematically advance the designer's goals over the user's goals \citep{conti2010malicious,gray2018darkpatterns,digeronimo2020uidark,mathur2021whatmakes,luguri2021shining}. \citet{mathur2019dark} documented widespread deceptive patterns that encourage users to choose actions that are easier or less effortful than the alternatives, while later work further clarified the design attributes, normative stakes, and user perceptions of such manipulative interfaces \citep{mathur2021whatmakes,bongard2021definitely}. In the realm of consent, \citet{nouwens2020dark} demonstrated that cookie banners obscure refusal options, prompt opt-out options that require multiple steps, or diminish attention through habituation. Subsequent work has connected consent-banner design to legal requirements, interaction criticism, and automated dark-pattern detection, showing that consent manipulation often emerges through multi-step flows, ambiguous labels, or asymmetric effort rather than through a single isolated screen \citep{gray2021legalrequirements,soe2022automated,hausner2021cookiebanners,kocyigit2023features}. \citet{utz2019informedconsent} further quantified how consent notice design can influence user comprehension of these notices and subsequent choices. Recent large-scale audits and cross-national usability studies echo and extend these results, highlighting common non-compliance patterns, mobile-specific frictions, evolving dark patterns, tracking consequences, and the continuing gap between nominal consent and practical control \citep{matte2020docookie,habib2022cookie,boumasims2023usukcmp,santos2021purpose,bielova2024designpatterns,degeling2019wevalue,machuletz2020multiple,demir2024interactiontools,tang2025consentchk,rasaii2025intractable,singh2026umbra}.

Adjacent work has also broadened how consent and privacy-choice interfaces are evaluated. Degeling et al.\ documented ecosystem-level changes in web privacy practices after the GDPR, while Singh et al.\ studied what kinds of cookie notices users prefer in the wild \citep{degeling2019wevalue,singh2022consentpreferences}. A parallel line of work on privacy notices and privacy labels has explored how complex data practices can be summarized into more usable formats, including privacy nutrition labels and mobile-app privacy labels \citep{kelley2009nutrition,kelley2010standardizing,kelley2013appdecision,li2022ioslabels,balash2024ioslabels,zhang2024expandablegrid}. This work is important for PSI because it shows that meaningful privacy choice depends not only on the existence of information, but on how that information and its associated actions are surfaced. Other work has examined adjacent forms of privacy control, such as deletion, opt-out mechanisms, app permissions, IoT privacy preferences, and mismatched privacy expectations, showing that meaningful alternatives are often hard to find, inconsistently labeled, or burdensome to exercise \citep{habib2019optoutchoices,habib2022choiceusability,kelley2013appdecision,rao2016mismatched,emaminaeini2017iotprivacy}. More recent analyses have also examined newer coercive models, such as accept-or-pay banners, and policy-facing reports have increasingly framed deceptive design as a consumer-protection and platform-accountability problem \citep{rasaii2023acceptorpay,oecd2022darkcommercial,luguri2021shining}. We build on this line of work, but focus on a different unit of analysis: the burden imposed along the pre-choice path before a meaningful non-accepting alternative becomes visible and actionable.

Decision-making under constraint is another recurring theme in this literature. Privacy choices are often made under repeated exposure, limited time, uneven consequences, and strong asymmetries in information and attention \citep{acquisti2015science}. At scale, the burden of policy reading makes fully informed notice-and-consent impractical in ordinary use \citep{mcdonald2008cost,wagner2023policies}, while experimental work shows that more usable privacy information can shape user decisions when it is presented at the right moment and in a usable format \citep{tsai2011privacyinfo,kelley2009nutrition,kelley2010standardizing}. Prior work on privacy notices and privacy-choice mechanisms also shows that effectiveness depends not only on legal content, but on how choices are surfaced, structured, and evaluated in practice \citep{schaub2015designspace,feng2021designspacechoices,habib2022choiceusability}. The warnings and permissions literature further shows that repeated prompts can degrade attention over time, encouraging users to progress through interfaces without sustained engagement \citep{sunshine2009ssl,egelman2009youvebeenwarned,felt2012android}. Related work on privacy framing and control likewise suggests that interface structure can shape both perceived agency and disclosure behavior even when underlying options remain formally available \citep{brandimarte2013misplaced,adjerid2013sleights}.

A third strand emphasized affect and ritual in UX. People use interfaces to enact values or identities, not strictly for goal directed behavior \citep{gray2018darkpatterns,fogg2003persuasive}. In the context of security and privacy, for example, security tasks like checking a lock icon or simply retyping a password provide affective security when the status has not changed \citep{bravo2013warnings}. We extend this analysis to consent flows, and think about a measurable ritual—performative scrolling—as an analytical lens on the \emph{privacy placebo} \citep{barth2017privacyparadox}. Recent work on cookie-banner design patterns further supports this lens by showing that defaults, presentation choices, and personalized banner designs can shape both consent behavior and perceived control \citep{grassl2021darkbright,biselli2024personalised,habib2021toggles}.

Finally, we focus on the structural gap between the appearance of available choice and the substance of meaningful control in consent interfaces. Interfaces can allow choice to pass as available while very deliberately making one path most salient and easy, while users can make reading appear diligent with their decision made anyway. When the structural gap on both sides align, the system accomplishes consent that is literally valid, but substantively thin \citep{solove2021paradox,bohme2010consentdialogs,norberg2007privacycalculus}. Our work differs from prior consent audits in two ways. First, rather than emphasizing prevalence, legal compliance, or final click outcomes, we focus on the \emph{pre-choice path} required before a meaningful alternative to ``Accept All'' becomes available. Second, rather than treating user dwell time or click counts as proxies for understanding, PSI isolates structural burden that can be reconstructed from first-encounter interface states and traversal logs. In that sense, PSI is intended as a complementary audit measure: it does not replace outcome-oriented or legal analyses, but adds a design-side account of whether alternatives are displaced offscreen, fragmented behind disclosure, or made fragile under keyboard traversal. It also complements recent work that studies privacy-choice usability, redesigned consent artifacts, and longitudinal audits of policy/interface change over time \citep{habib2022choiceusability,pearman2022hipaa,10.1145/3772363.3798570}.

\section{Audit Framework}

\subsection{Signals and Index Definition}
We instrument consent flows in a minimal, practical way. A scripted \emph{least-effort} agent opens the banner and follows the shortest visible path to the first \emph{meaningful} control (Reject/Customize/Save). Along that path we record only five observable transitions in plain terms—\texttt{scroll}, \texttt{expand}, \texttt{toggle}, \texttt{focus-loop}, and \texttt{action}—derived from the DOM and timing\citep{messick1995validity,borsboom2004validity,munzner2009nested}.

From the event timeline we compute four quantities that matter before a meaningful choice becomes actionable. \(D\) is the cumulative pixels scrolled; normalizing by the viewport height \(\mathrm{vh}\) gives distance in “viewports.” \(T\) is time-to-primary under deterministic animation timing. \(F\) counts \texttt{focus-loop} occurrences (e.g., tab order that cycles without progress). \(H\) counts hidden-but-material toggles that had to be revealed on the way to the first meaningful control. These are summarized by the Performative Scrolling Index:
\[
\mathrm{PSI}=\alpha\cdot\frac{D}{\mathrm{vh}}+\beta\cdot T+\gamma\cdot F+\delta\cdot H,
\]
with profile weights \((\alpha,\beta,\gamma,\delta)\) chosen by the auditor (e.g., accessibility may up-weight \(\gamma\)). We report PSI alongside compact companion signals: time-to-primary (seconds), distance-to-choice (in viewports), a \emph{granularity-exposed} flag indicating whether a meaningful alternative to “Accept All” is visible on the first pane, and a \emph{reversibility} flag indicating a persistent, one-click path to revisit or revoke consent.

\subsubsection{Why these components.}
The four components are not intended to exhaust every possible form of consent friction. Following construct-validation and composite-indicator guidance, we define the measurement target narrowly: observable pre-choice burden before a meaningful alternative becomes visible and actionable \citep{cronbach1955construct,messick1995validity,borsboom2004validity,nardo2008compositeindicators}. The components correspond to four observable ways that access to such an alternative can be delayed in an interface audit: spatial displacement (\(D/\mathrm{vh}\)), temporal delay (\(T\)), focus-order fragility (\(F\)), and structural concealment (\(H\)). This makes the component set sufficient for the scoped question PSI asks---how much pre-choice interaction burden is imposed before a meaningful alternative becomes visible and actionable---while leaving semantic manipulation, affective reassurance, and legal adequacy to complementary analyses.

\subsubsection{Why additive aggregation.}
We use an additive form because each component represents a distinct and monotonically increasing source of pre-choice burden: more scrolling, longer delay, more focus loops, or more hidden reveals should not reduce the audit score. Additivity also keeps the score decomposable: every PSI value can be traced back to the contribution of each component. We do not claim that the additive form is a psychological model of how users experience burden. Instead, it is a transparent audit convention whose usefulness is evaluated through component reporting, robustness checks, comparison against simpler baselines, and sensitivity to alternative weighting profiles \citep{messick1995validity,munzner2009nested,nardo2008compositeindicators}.

\subsubsection{Component-first interpretation.}
Although PSI is reported as a scalar summary, the primary measurement output of the audit is the component vector
\((D/\mathrm{vh}, T, F, H)\). We use the scalar PSI to support compact comparison across interfaces, but we do not treat any single weighting profile as uniquely correct. Instead, aggregate PSI should be interpreted together with its component decomposition and with sensitivity analysis over plausible weighting profiles. This component-first interpretation reduces dependence on any one normative choice about how distance, delay, focus fragility, and hidden disclosure should be traded off.

PSI is intended as a transparent composite index rather than a latent psychological scale. Accordingly, the weights $(\alpha,\beta,\gamma,\delta)$ should be interpreted as \emph{audit priorities} that determine how much an evaluator wishes to penalize four observable sources of pre-choice burden: traversal distance, delay, keyboard fragility, and hidden-but-material disclosure. This choice is deliberate. At the current stage, we do not claim that a single canonical weighting is ``true'' across all accessibility profiles, devices, or policy settings. Instead, PSI is designed to expose its assumptions directly, so that auditors can report both the chosen profile and sensitivity of rankings to plausible alternatives.

\subsubsection{Validation scope.}
Our validation is structural rather than behavioral. We do not claim in this paper that PSI predicts comprehension, satisfaction, legal sufficiency, or downstream welfare. Instead, we evaluate whether PSI behaves as a useful audit construct: whether it orders known interface patterns in expected ways, remains stable across plausible weighting profiles, distinguishes burden sources that simpler heuristics collapse, and exposes traversal-policy differences that would otherwise be missed. This follows a layered view of validation: the present paper validates the audit abstraction and implementation behavior, while leaving direct user-outcome validation to future controlled studies \citep{messick1995validity,munzner2009nested}.

\noindent\textbf{Default profile.}
Unless stated otherwise, we report an equal-weight profile
\((\alpha,\beta,\gamma,\delta)=(1,1,1,1)\) as a transparent baseline. We use this profile not because the four components are assumed to have identical psychological effects, but because equal weighting provides a neutral first-pass summary in which each observable burden source contributes monotonically to the total score. To avoid over-interpreting this scalar, all deployment analyses also report component summaries and robustness checks across alternative weighting profiles.

\noindent\textbf{Illustrative alternative profiles.}
To make the weighting scheme more concrete, we consider three additional profiles beyond the equal-weight baseline $(1,1,1,1)$. An \emph{accessibility-oriented} profile $(1,1,2,1)$ places greater emphasis on focus fragility and keyboard burden, making stalled or looping keyboard paths more costly. A \emph{delay-sensitive} profile $(1,2,1,1)$ places greater emphasis on temporal friction, including animation-gated or staged interactions that slow access to meaningful alternatives. A \emph{disclosure-sensitive} profile $(1,1,1,2)$ places greater emphasis on hidden-but-material reveals, making collapsed settings panels or buried refusal paths more costly. We do not claim that any one of these profiles is universally correct; rather, they illustrate how PSI can be adapted to distinct audit priorities while keeping those assumptions explicit and reviewable.

\begin{table}[t]
\centering
\footnotesize
\caption{Illustrative PSI weighting profiles. Higher weights place greater emphasis on the corresponding burden source.}
\label{tab:profiles}
\setlength{\tabcolsep}{4.5pt}
\begin{tabular}{@{}>{\raggedright\arraybackslash}p{2.35cm}cccc>{\raggedright\arraybackslash}p{3.15cm}@{}}
\toprule
\rowcolor{TableHeader}
\textbf{Profile} & \textbf{$\alpha$} & \textbf{$\beta$} & \textbf{$\gamma$} & \textbf{$\delta$} & \textbf{Primary emphasis} \\
\midrule
Default baseline       & 1 & 1 & 1 & 1 & Neutral first-pass audit of overall pre-choice burden \\
\rowcolor{TableStripe}
Accessibility-oriented & 1 & 1 & 2 & 1 & Greater penalty for keyboard fragility and focus-loop burden \\
Delay-sensitive        & 1 & 2 & 1 & 1 & Greater penalty for temporal friction and staged delay \\
\rowcolor{TableStripe}
Disclosure-sensitive   & 1 & 1 & 1 & 2 & Greater penalty for concealed controls and hidden reveals \\
\bottomrule
\end{tabular}
\end{table}

Reported PSI values should therefore always be interpreted together with the profile under which they were computed, since different profiles foreground different sources of burden.

\begin{figure}[t]
    \centering
    \resizebox{0.85\columnwidth}{!}{%
    \begin{tikzpicture}[
        node distance=0.8cm and 1.2cm,
        box/.style={rectangle, rounded corners, draw, thick, text width=3.5cm, align=center, minimum height=1cm, font=\sffamily\footnotesize},
        obstructive/.style={box, fill=red!5, draw=red!60!black},
        honest/.style={box, fill=green!5, draw=green!60!black},
        decision/.style={diamond, aspect=2, draw, thick, fill=red!5, draw=red!60!black, text width=2cm, align=center, font=\sffamily\footnotesize, inner sep=0pt},
        arrow/.style={-{Latex[scale=1]}, thick, draw=gray!80},
        dashedarrow/.style={-{Latex[scale=1]}, thick, dashed, draw=gray!80}
    ]

    \node[obstructive] (obs) {\textbf{Obstructive Design}};
    \node[obstructive, below=of obs] (highb) {High Performative Burden (PSI) \\ \& Low Comprehension (CSI)};
    \node[obstructive, below=of highb] (fatigue) {User Frustration \& Fatigue};
    \node[decision, below=of fatigue] (choice) {Make a Choice};
    
    \node[below=0.5cm of choice, font=\sffamily\footnotesize, text=gray!80!black] (ritual) {Ritualistic Acceptance};
    \node[obstructive, below=0.1cm of ritual] (unref) {\textbf{Unreflective Acceptance}};

    \node[honest, left=1cm of highb] (hon) {\textbf{Honest Design}};
    \node[honest, below=of hon] (lowb) {Low Performative Burden (PSI) \\ \& High Comprehension (CSI)};
    \node[honest, below=of lowb] (informed) {\textbf{Informed, Reflective Decision}};

    \begin{scope}[on background layer]
        \node[rectangle, draw=green!50!black, fill=green!2, thick, dashed, inner sep=0.3cm, 
              label={[font=\sffamily\bfseries\footnotesize, text=green!50!black]above:Break The Cycle: Apply Honest Design}, 
              fit=(hon) (lowb) (informed)] (intervention_box) {};
    \end{scope}

    \draw[arrow] (obs) -- (highb);
    \draw[arrow] (highb) -- (fatigue);
    \draw[arrow] (fatigue) -- (choice);
    \draw[arrow] (choice) -- (ritual);
    \draw[arrow] (ritual) -- (unref);
    
    \draw[dashedarrow] (unref.east) -- ++(0.8,0) |- node[pos=0.25, right, font=\sffamily\footnotesize, text=gray!80!black] {Reinforces Loop} (obs.east);

    \draw[arrow] (obs.west) -| node[above, xshift=1cm, font=\sffamily\footnotesize, text=gray!80!black] {Intervention} (hon.north);
    \draw[arrow] (hon) -- (lowb);
    \draw[arrow] (lowb) -- (informed);

    \end{tikzpicture}%
    } 
    \caption{Conceptual logic of the privacy-placebo framing. Structural features of obstructive consent interfaces can increase pre-choice burden and reduce comprehension support, encouraging routine or low-reflection interaction. Intervening with more supportive design can lower burden and improve the conditions for informed choice.}
    \label{fig:consent_loop}
\end{figure}
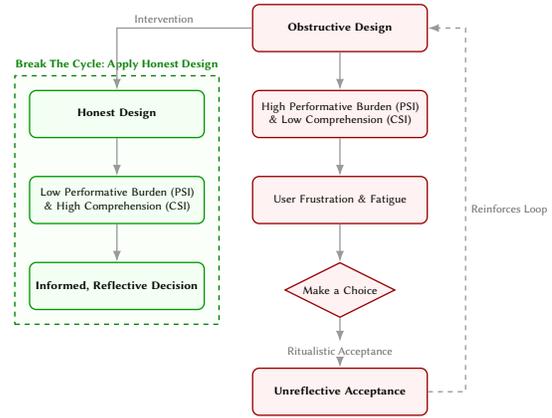

Figure~\ref{fig:consent_loop} summarizes the conceptual logic motivating PSI. The figure should not be read as a validated behavioral model; rather, it illustrates the design-side pathway through which consent-surface structure can raise pre-choice burden before a meaningful non-accepting alternative becomes visible and actionable.

In this framing, PSI captures the burden imposed by the structure of the interface, while companion signals such as assurance cues and comprehension support help interpret whether a design supports meaningful control or merely its appearance. The goal is therefore not simply to reduce interaction effort in the abstract, but to reduce avoidable pre-choice burden while preserving visible alternatives, local explanation, and reversibility.

In practice, PSI rises when alternatives are displaced offscreen (\(D/\mathrm{vh}\uparrow\)), fragmented behind disclosure (\(H\uparrow\)), or when staged modals reset focus between panes (\(F\uparrow\)). Lowering PSI while raising CSI—by co-presenting alternatives, keeping explanations local to actions, and making reversibility persistent—moves designs away from placebo-prone profiles without constraining legitimate legal text.

\subsection{Operationalization and Audit Protocol}

Our emphasis on keyboard traversal, locality, and co-present alternatives aligns with widely adopted accessibility and UX guidance, as well as prior frameworks for evaluating privacy notices and privacy-choice usability. WCAG 2.1 requires a meaningful \emph{focus order} and operable keyboard paths \citep{schaub2015designspace,habib2022choiceusability,wcag21,waiFocusOrder,nngAccordion2023,idfProgressiveDisclosure,bollinger2022automating}; progressive disclosure should not bury primary choices or sever explanations from controls \citep{nngAccordion2023,idfProgressiveDisclosure}. Focusing on interface-derived signals also supports automation and repeated audits, complementing recent efforts to assess consent interfaces at scale \citep{bollinger2022automating}.

In order to achieve audits that are reproducible across different sites and devices, we use conservative defaults in practice to operationalize PSI \citep{habib2022cookie,boumasims2023usukcmp,matte2020docookie}. We define a distance when the first paint of the consent surface is reached to the first frame where a materially different, non-accept option is not only in the DOM but also visible inside of the viewport and reachable with a single primary interaction \citep{nngAccordion2023,idfProgressiveDisclosure}. Time is defined as the scripted interaction plus any animation delays denoted by computed style; when activation is gated—even if due to easings or debounced listeners that are purely decorative—we treat these as time costs because such “micro-frictions” alter the choice architecture \citep{johnson2012choicearchitecture}. Focus loops are only included when tab order or roving \texttt{tabindex} has cycled through controls without any meaningful control being advanced; when focus moves due to incidental focus shifts that are legitimate disclosures (e.g., expanding a details element that lands the focus inside of it), we do not fault this if it is notable that the non-accept control is co-present \citep{wcag21,waiFocusOrder}. Hidden-but-material controls (e.g., refusal, category-level controls) are noted when they are occluded behind collapsed panels, drawers, or steppers, such that an action revealing the non-accepting option is needed before it is accessible \citep{nngAccordion2023,idfProgressiveDisclosure,mathur2019dark}.

\paragraph{Meaningful alternative.}
We define a \emph{meaningful alternative} as the first non-accepting control that is both \emph{visible} and \emph{actionable} at first encounter. Concretely, a control qualifies if it satisfies all of the following: (i) it corresponds to a non-accept path such as \emph{Reject}, \emph{Customize}, \emph{Manage settings}, or \emph{Save choices}; (ii) it is rendered within the current viewport without requiring further scroll to discover it; (iii) it is reachable with one primary interaction under the current traversal policy; and (iv) activating it advances the user toward refusing, narrowing, or revising consent rather than merely presenting the accepting path again. Controls that are present in the DOM but visually occluded, collapsed behind disclosure, detached from keyboard reachability, or too euphemistic to establish a non-accept path are not counted as meaningful alternatives at first encounter.

\paragraph{Decision rules for ambiguous controls.}
Because many consent surfaces use euphemistic or indirect labels, we apply conservative decision rules when identifying the first meaningful alternative. Labels such as ``Manage experience,'' ``Learn more,'' ``Privacy choices,'' or ``More options'' are counted as meaningful only if activating them clearly advances the user toward refusing, narrowing, or revising consent. A control that merely reveals explanatory text, repeats the accepting path, or leads to a general privacy-policy page is not counted as the first meaningful alternative. Likewise, a control that appears in the DOM or is partially visible on screen is not sufficient: it must be visible within the effective consent-surface viewport, enabled, not occluded, reachable under the declared traversal policy, and actionable through one primary interaction. This conservative treatment of ambiguous labels also reflects broader evidence that privacy-policy and privacy-control language can be difficult to parse automatically or by users, motivating structured decision rules and evidence frames rather than purely semantic inference \citep{reidenberg2016ambiguity,harkous2018polisis,harkous2016pribots}.

We apply the same logic to route selection. Expanding a disclosure element does not by itself satisfy the audit objective unless the expansion reveals a control that materially advances refusal, customization, or saving narrowed preferences. If multiple non-accepting alternatives are visible on the first pane, we follow the shortest visible route to the most direct actionable path, privileging an immediately available \emph{Reject} or \emph{Save choices} control over a longer \emph{Customize} route unless customization is the only route that enables non-accepting choice. Focus movement is counted as a focus loop only when traversal cycles through controls without progress toward a meaningful alternative; legitimate focus relocation caused by disclosure is not penalized if it moves focus closer to a visible and actionable non-accepting control. When the consent surface itself scrolls independently from the page, distance is computed relative to the effective consent-surface viewport.

\paragraph{Audit procedure.}
For each consent surface, the audit proceeds in five steps. First, we capture the first-encounter state at a fixed breakpoint and record the initial viewport. Second, under a declared traversal policy (pointer-only or keyboard-only), we follow the shortest visible route to the first meaningful alternative. Third, along that route we log only five event classes: \texttt{scroll}, \texttt{expand}, \texttt{toggle}, \texttt{focus-loop}, and \texttt{action}. Fourth, we compute the PSI components $D/\mathrm{vh}$, $T$, $F$, and $H$, and report PSI together with companion audit signals including time-to-primary, distance-to-choice, granularity-exposed, and reversibility. Fifth, we export an annotated first-encounter strip marking the first frame at which a non-accepting alternative becomes both visible and actionable. This output is intended to make the resulting score inspectable by reviewers rather than opaque.

\begin{figure}[htpb]
    \centering
    \resizebox{\columnwidth}{!}{%
    \begin{tikzpicture}[
        node distance=0.4cm and 0.6cm,
        box/.style={rectangle, rounded corners, draw, thick, text width=2.6cm, align=center, minimum height=1.1cm, font=\sffamily\scriptsize},
        phase_label/.style={font=\sffamily\bfseries\scriptsize, text=gray!80!black, anchor=south},
        arrow/.style={-{Latex[scale=0.9]}, thick, draw=gray!80},
        init/.style={fill=blue!5, draw=blue!60!black},
        traverse/.style={fill=orange!5, draw=orange!60!black},
        compute/.style={fill=purple!5, draw=purple!60!black},
        output/.style={fill=green!5, draw=green!60!black}
    ]

    \node[box, init] (state) {\textbf{First-Encounter} \\ State (Fixed Breakpoint)};

    \node[box, traverse, right=0.8cm of state, yshift=0.8cm] (policy) {\textbf{Traversal Policy} \\ Pointer or Keyboard};
    \node[box, traverse, below=0.3cm of policy] (route) {\textbf{Shortest Route} \\ to Meaningful Alt.};
    \draw[arrow] (policy) -- (route);

    \node[box, compute, right=0.8cm of policy] (events) {\textbf{Event Trace} \\ Scroll, Expand, Toggle, Focus, Action};
    \node[box, compute, below=0.3cm of events] (components) {\textbf{PSI Components} \\ Norm. Dist., Time, Focus, Reveals};
    \draw[arrow] (events) -- (components);

    \node[box, output, right=0.8cm of events] (score) {\textbf{Aggregated PSI} \\ \& Companion Signals};
    \node[box, output, below=0.3cm of score] (evidence) {\textbf{Evidence Frame} \\ (Annotated Action)};

    \draw[arrow] (state.east) -- ++(0.4,0) |- (policy.west);
    
    \draw[arrow] (route.east) -- ++(0.4,0) |- (events.west);
    
    \draw[arrow] (components.east) -- ++(0.4,0) |- (score.west);
    \draw[arrow] (components.east) -- ++(0.4,0) |- (evidence.west);

    \begin{scope}[on background layer]
        \node[rectangle, draw=blue!40, fill=blue!2, thick, dashed, inner sep=0.25cm, 
              fit=(state)] (bg_init) {};
        \node[phase_label] at (bg_init.north) {1. Init};

        \node[rectangle, draw=orange!40, fill=orange!2, thick, dashed, inner sep=0.25cm, 
              fit=(policy) (route)] (bg_traverse) {};
        \node[phase_label] at (bg_traverse.north) {2. Audit};

        \node[rectangle, draw=purple!40, fill=purple!2, thick, dashed, inner sep=0.25cm, 
              fit=(events) (components)] (bg_compute) {};
        \node[phase_label] at (bg_compute.north) {3. Compute};

        \node[rectangle, draw=green!40, fill=green!2, thick, dashed, inner sep=0.25cm, 
              fit=(score) (evidence)] (bg_output) {};
        \node[phase_label] at (bg_output.north) {4. Output};
    \end{scope}

    \end{tikzpicture}%
    } 
    \caption{The full PSI audit pipeline. The workflow compresses the first-encounter state into an observable trace via a reproducible traversal, culminating in an aggregated score and inspectable evidence.}
    \label{fig:psi_workflow}
\end{figure}
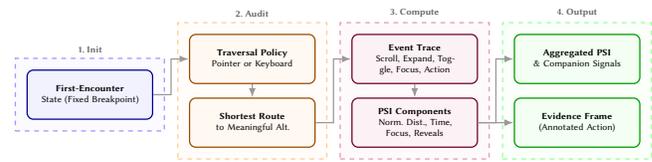

Figure~\ref{fig:psi_workflow} summarizes the end-to-end audit pipeline, from first-encounter capture and traversal-policy selection to event logging, PSI computation, and inspectable output.

The workflow makes explicit that PSI is not just a scalar score: it is derived from a compact interaction trace and paired with annotated evidence showing when a meaningful alternative first becomes visible and actionable.

\paragraph{Implementation details.}
We implemented the audit prototype as a browser-driven instrumentation script running in a Chromium-based environment at fixed desktop and mobile breakpoints (desktop: 1440$\times$900; mobile: 390$\times$844). For each surface, the auditor records the first stable consent state after page load and then executes the declared traversal policy (pointer-only or keyboard-only) under deterministic timing. Animation costs are derived from computed style when available; otherwise, we apply a conservative fixed wait budget only for transitions that visibly gate access to the next actionable state. We count a \emph{primary interaction} as a single click, keypress, or toggle action that materially advances the path toward a non-accepting alternative. Dynamically loaded CMP elements are included only after the consent surface reaches a stable rendered state or after a bounded wait window, after which the audit records the earliest actionable configuration observed. These implementation choices are intended to make PSI reproducible across auditors while avoiding dependence on user-specific dwell or unconstrained exploration.

To ensure that findings are portable, we fix viewport height at a common breakpoint and report sensitivity of results to $\pm 20\%$ changes; the qualitative ordering of patterns remains consistent under these conditions, mitigating single–device-profile artifacts \citep{habib2022cookie,boumasims2023usukcmp}. Audits are run twice: a pointer-only policy that follows the shortest visible path and a keyboard-only policy that respects ARIA roles and the page’s declared \texttt{tabindex} \citep{wcag21,waiFocusOrder}. A design that “passes” under pointer but stalls under keyboard is flagged as fragile; the accompanying report lists minimal edits to align the paths and includes an annotated strip of screenshots marking the first frame where non-accept options become visible, supporting automation and reviewer interpretability \citep{bollinger2022automating,redmiles2019privacystudy}.

\section{Feasibility Deployment and Illustrative Results}
We report four kinds of evidence. First, we analyze canonical consent patterns---\emph{Scroll Wall}, \emph{Accordion Disclosure}, \emph{Multi-step Modal}, and \emph{Co-present Choice}---to illustrate how PSI changes as alternatives are displaced, fragmented, gated, or immediately available. Second, we apply PSI in a live deployment across 500 websites, auditing each site under desktop and mobile breakpoints and under pointer and keyboard traversal policies. Third, we compare PSI against simpler baseline heuristics such as click count, first-pane visibility, distance-to-choice, and keyboard reachability. Fourth, we report detector reliability, inter-rater reproducibility, weighting sensitivity, and rendering sensitivity. The goal of this section is not to estimate web-wide prevalence, but to show that PSI can be computed reproducibly at scale and that it captures burden patterns that simpler checklist-style measures can miss.

\subsection{Canonical patterns and measurement setup}
We present three higher-burden canonical embodiments---\emph{Scroll Wall}, \emph{Accordion Disclosure}, and \emph{Multi-step Modal}---and use \emph{Co-present Choice} as a low-burden comparison condition. Each condition uses the same policy skeleton and material choices; only the layout and interaction structure differ. A least-effort policy takes the shortest visible route to the first meaningful control (Reject/Customize/Save) and logs \texttt{scroll}, \texttt{expand}, \texttt{toggle}, \texttt{focus-loop}, and \texttt{action} events to ascertain distance-to-choice for viewports, time-to-primary for fixed animation costs, counts of focus loops, and counts of hidden-but-material toggles. This arrangement purposely decouples results from individual differences, and avoids interpreting dwell or scroll counts as a stand-ins for understanding \citep{sunshine2009ssl,egelman2009youvebeenwarned,felt2012android}. Ambiguity in wording of the policy and clarity of disclosure are treated as structural confounds and documented via checks of locality and co-presence \citep{reidenberg2016ambiguity,korunovska2020comprehension,chen2021fogccpa}. Finally, choice architecture and accessibility guidance provide motivation for our co-present alternative, locality of explanations, and keyboard operability checks \citep{johnson2012choicearchitecture,wcag21,waiFocusOrder,nngAccordion2023,idfProgressiveDisclosure}. These canonical patterns serve as controlled design archetypes. We use them to separate structural effects from site-specific copy, branding, or vendor-specific implementation details before turning to the live deployment. We selected these three patterns because each isolates a common source of pre-choice burden: offscreen displacement in the Scroll Wall, disclosure fragmentation in the Accordion pattern, and stepwise gating plus focus fragility in the Multi-step Modal.

Taken together, these archetypes are useful not only for controlled comparison, but also as redesign templates: each isolates a structural burden source that can be reduced through relatively direct interface changes.

\subsection{PSI bounds and invariants}
Because PSI is additive over distance, time, focus loops, and hidden toggles, several design-side facts follow directly:

\emph{Co-presence lower bound.} If a meaningful alternative is visible on the first pane and reachable with one primary interaction, then \(H{=}0\) and \(D/\mathrm{vh}\le 1\), implying \(\mathrm{PSI}\ge \beta T\) with \(\alpha(D/\mathrm{vh})\) minimized. This is the design-side lower bound for any fixed \((\alpha,\beta,\gamma,\delta)\) profile and aligns with evidence that co-present choices improve comprehension affordances without suppressing completion \citep{utz2019informedconsent,habib2022cookie}.

\emph{Fragmentation penalty.} Splitting alternatives across panes or folded disclosures raises \(H\) (reveals) and often \(F\) (focus resets), increasing PSI even when legal text is unchanged. This reproduces common failure modes observed in CMP evaluations and enforcement complaints \citep{nouwens2020dark,noyb2024cookiereport}.

\emph{Default salience vs.\ locality tension.} Increasing button salience and progress cues can raise assurance (AAI) without improving locality or co-presence, producing positive divergence \( \mathrm{DIV}{=}\mathrm{AAI}{-}\mathrm{CSI} \) \citep{gray2018darkpatterns,thaler2008nudge}. Reviews consistently show that celebratory microcopy and dominant “Accept” affordances co-occur with displaced alternatives \citep{matte2020docookie,habib2022cookie}.

\emph{Accessibility invariant.} If keyboard traversal induces a focus loop before a meaningful control, then any pointer-path “pass” is fragile; PSI must strictly increase under a keyboard-only policy (\(F\uparrow, T\uparrow\)) per WCAG 2.1 SC 2.4.3/2.1.1 \citep{wcag21,waiFocusOrder}. Robust designs therefore minimize PSI under both policies.

These invariants let auditors argue from structure rather than samples of behavior: for a given flow, moving a meaningful alternative above the fold and co-present with “Accept All” strictly reduces the distance term, and replacing nested panes with local toggles strictly reduces \(H\) (and often \(F\))—independent of who will eventually use the interface \citep{johnson2012choicearchitecture,idfProgressiveDisclosure}.

\subsection{Worked example: first-encounter traversal}
To make the computation concrete, we summarize one field-audit vignette. On an anonymized CMP deployed on a top-ranked news site at a desktop breakpoint, the least-effort traversal produced the event strip
{\small \texttt{EV\_SCROLL} $\rightarrow$ \texttt{EV\_EXPAND} $\rightarrow$ \texttt{EV\_TOGGLE} $\rightarrow$ \texttt{EV\_ACTION}}.
In this case, the first non-accept alternative became available only after viewport traversal and a collapsed-panel reveal. Under our framework, this path increases both the distance term and the hidden-reveal term relative to a co-present design, even though the legal text itself is unchanged.

The same vignette also illustrates why parity across traversal policies matters. Under pointer traversal, the shortest visible route reached a non-accept option after scroll and disclosure. Under keyboard traversal, the same interface introduced a focus loop before the non-accepting path became reachable, strictly increasing burden under the keyboard policy. This is exactly the kind of design fragility PSI is intended to expose: not merely that the interface contains a refusal path somewhere in the DOM, but that access to that path is displaced or fragile at first encounter.

\subsection{Live deployment}

Table~\ref{tab:deployment-summary} summarizes component medians and PSI distributions across device and traversal conditions. Across the deployment, mobile layouts exhibited higher median burden than desktop layouts, and keyboard traversal increased PSI relative to pointer traversal under both breakpoints. The largest difference appeared under mobile keyboard traversal, where smaller viewports and fragile focus behavior compounded the burden imposed before a meaningful non-accepting alternative became visible and actionable.

Of the 2,000 possible site-condition traces, 1,327 produced successful first-encounter consent-surface traces included in PSI aggregation. Table~\ref{tab:trace-accounting} summarizes how all attempted traces were handled. We report this accounting separately from PSI results because failures are themselves informative about deployment feasibility, but should not be silently folded into scalar PSI scores. Failed, absent, blocked, timed-out, and unrecoverable traversal-failure traces were retained in the audit log and excluded from aggregate PSI scoring. Traversal failures after a stable consent state were analyzed separately as fragility evidence rather than folded into the scalar PSI distributions below.

\begin{table}[t]
\centering
\footnotesize
\caption{Trace accounting for the live deployment. Counts are reported over 500 websites $\times$ 2 device breakpoints $\times$ 2 traversal policies, yielding 2,000 possible site-condition traces.}
\label{tab:trace-accounting}
\setlength{\tabcolsep}{4.5pt}
\renewcommand{\arraystretch}{1.16}
\begin{tabular}{@{}p{3.05cm}r p{3.35cm}@{}}
\toprule
\rowcolor{TraceHeader}
\textbf{Trace outcome} & \textbf{Count} & \textbf{Treatment in analysis} \\
\midrule
\rowcolor{TraceSuccess}
Successful first-encounter trace & \textbf{1,327} & \included{} in PSI aggregation \\
No consent surface detected & 284 & \excluded{} from PSI; retained in audit log \\
\rowcolor{TraceFail}
Blocked or inaccessible CMP iframe & 126 & \excluded{} from PSI; retained as deployment failure \\
Timeout before stable consent state & 104 & \excluded{} from PSI; retained as timing failure \\
\rowcolor{TraceWarn}
Ambiguous non-accepting alternative & 71 & \flagged{} for adjudication; excluded unless resolved \\
Traversal failure after stable state & 58 & \flagged{} separately; excluded from aggregate PSI \\
\rowcolor{TraceFail}
Rendering or instrumentation error & 30 & \excluded{} from PSI; retained as tooling failure \\
\midrule
\rowcolor{TraceTotal}
\textbf{Total attempted traces} & \textbf{2,000} & \textbf{Full deployment denominator} \\
\bottomrule
\end{tabular}
\end{table}

This accounting also clarifies the denominator for the deployment claims. The PSI distributions below describe successful, inspectable first-encounter traces rather than all visited websites. We therefore interpret the deployment as feasibility and distributional evidence, not as a prevalence estimate of all consent interfaces on the web.

\begin{table}[t]
\centering
\footnotesize
\caption{Deployment summary across device and traversal conditions. Component entries report marginal medians; Median PSI is computed from each trace's scalar PSI and therefore need not equal the sum of the displayed component medians. Highlighted cells mark the highest observed median and P90 burden across conditions.}
\label{tab:deployment-summary}
\setlength{\tabcolsep}{4.2pt}
\renewcommand{\arraystretch}{1.15}
\begin{tabular}{@{}lccccccc@{}}
\toprule
\rowcolor{TableHeader}
\textbf{Condition} & \textbf{Traces} & \textbf{$D/\mathrm{vh}$} & \textbf{$T$} & \textbf{$F$} & \textbf{$H$} & \textbf{Median PSI} & \textbf{P90 PSI} \\
\midrule
Desktop, pointer  & 342 & 0.8 & 0.7 & 0.0 & 1.0 & 2.1 & 4.8 \\
\rowcolor{TableStripe}
Desktop, keyboard & 329 & 0.9 & 0.8 & \midcell{0.5} & 1.0 & 2.7 & 5.6 \\
Mobile, pointer   & 337 & \midcell{1.2} & 0.8 & 0.0 & 1.0 & 2.8 & 6.1 \\
\rowcolor{PsiHigh}
Mobile, keyboard  & 319 & \textbf{1.4} & \textbf{0.9} & \textbf{0.7} & 1.0 & \textbf{3.5} & \textbf{7.0} \\
\bottomrule
\end{tabular}
\end{table}

We also categorized audited surfaces by the dominant structural pattern observed at first encounter. Table~\ref{tab:deployment-archetypes} reports median PSI by archetype, aggregating across device and traversal conditions. Co-present designs had the lowest median burden because a meaningful non-accepting alternative was visible and actionable immediately. Scroll Wall designs increased PSI primarily through distance-to-choice. Accordion designs increased PSI through hidden-reveal burden, even when pixel distance was relatively low. Multi-step Modal designs produced the highest burden because they combined staged delay, hidden structure, and keyboard fragility.

\begin{table}[t]
\centering
\footnotesize
\caption{PSI by dominant consent-surface archetype in the live deployment. Shading in the Median PSI column marks lower, moderate, and higher burden.}
\label{tab:deployment-archetypes}
\setlength{\tabcolsep}{4.4pt}
\renewcommand{\arraystretch}{1.15}
\begin{tabular}{@{}lcccc@{}}
\toprule
\rowcolor{TableHeader}
\textbf{Archetype} & \textbf{Sites} & \textbf{Median PSI} & \textbf{IQR} & \textbf{Primary driver} \\
\midrule
Co-present Choice & 84 & \bestcell{1.1} & 0.8--1.6 & Immediate alternative \\
\rowcolor{TableStripe}
Scroll Wall       & 102 & \midcell{2.7} & 2.0--3.8 & Distance \\
Accordion         & 113 & \midcell{3.4} & 2.5--4.6 & Hidden reveals \\
\rowcolor{PsiHigh}
Multi-step Modal  & 96 & \textbf{4.2} & \textbf{3.1--5.7} & Staged gating \\
Other/Mixed       & 105 & \midcell{3.0} & 2.0--4.4 & Mixed structure \\
\bottomrule
\end{tabular}
\end{table}

These results should be interpreted as distributional feasibility evidence rather than as definitive prevalence estimates. Even so, the deployment strengthens the central claim of the paper: PSI scales beyond isolated examples, produces interpretable burden distributions, and exposes systematic differences across device and traversal conditions.

\subsection{Traversal-policy effects}
Because each site was audited under both pointer and keyboard traversal, we can compare how much additional burden appears when users must rely on focus order rather than direct pointer interaction. As shown in Table~\ref{tab:deployment-summary}, keyboard traversal increased median PSI under both desktop and mobile conditions. This difference is substantively important: an interface may expose a meaningful alternative somewhere in the DOM while still making that alternative fragile or delayed under keyboard traversal. We therefore treat traversal policy as part of the audit condition rather than as an implementation detail.

\subsection{Structural validation and baseline comparison}

Because PSI is intended as a design-side audit metric, we evaluate its structural validity using the deployment data rather than claiming direct behavioral validation. We use three checks. First, we test \emph{known-groups validity}: whether PSI orders consent-surface archetypes in ways predicted by the framework. Second, we compare PSI against simpler baseline heuristics, including click count, first-pane visibility, distance-to-choice only, hidden-panel count, and keyboard reachability. Third, we examine whether the component vector \((D/\mathrm{vh}, T, F, H)\) adds diagnostic information beyond any single baseline.

\paragraph{Known-groups validity.}
If PSI is measuring structural pre-choice burden, then designs that expose a meaningful alternative immediately should score lowest, while designs that fragment alternatives across staged panes or fragile focus paths should score highest. This pattern appears in the deployment: Co-present Choice has the lowest median PSI, while Multi-step Modal has the highest median PSI. Scroll Wall and Accordion designs occupy the middle but differ in their component profiles, with Scroll Wall burden driven primarily by distance and Accordion burden driven primarily by hidden reveals.

\paragraph{Baseline comparison.}
Table~\ref{tab:baseline-comparison} summarizes representative cases where PSI provides more diagnostic information than simpler checklist-style metrics. Click count alone fails to distinguish scroll-heavy and disclosure-heavy paths with similar action counts. A binary first-pane visibility measure identifies the clearest low-burden designs but misses staged gating and keyboard fragility. Distance-to-choice captures displacement but misses hidden-reveal burden. Keyboard reachability flags accessibility failure but does not quantify how much burden is imposed before the alternative becomes actionable \citep{messick1995validity,munzner2009nested,schaub2015designspace,habib2022choiceusability}.

\begin{table*}[t]
\centering
\footnotesize
\caption{Baseline comparison. PSI adds diagnostic value by decomposing burden into distance, time, focus fragility, and hidden reveals rather than relying on a single checklist-style heuristic.}
\label{tab:baseline-comparison}
\setlength{\tabcolsep}{4.5pt}
\renewcommand{\arraystretch}{1.18}
\begin{tabular}{@{}p{2.2cm}cccc>{\columncolor{TableSummary}}p{4.2cm}@{}}
\toprule
\rowcolor{TableHeader}
\textbf{Case} & \textbf{Clicks} & \textbf{First pane?} & \textbf{Keyboard?} & \textbf{PSI} & \textbf{What PSI reveals} \\
\midrule
Co-present Choice & 1 & Yes & Yes & \bestcell{1.1} & Low burden because the alternative is immediately visible and actionable. \\
\rowcolor{TableStripe}
Scroll Wall & 2 & No & Yes & 2.7 & Similar click count to some disclosure paths, but burden comes from viewport displacement. \\
Accordion Disclosure & 2 & No & Yes & 3.4 & Click count is low, but PSI exposes hidden-reveal burden. \\
\rowcolor{TableStripe}
Multi-step Modal & 3 & No & Partial & \highcell{4.2} & Burden comes from staged gating plus focus fragility, not just number of clicks. \\
Keyboard-fragile banner & 1--2 & Yes & Fragile & \midcell{3.1} & Pointer path appears acceptable, but keyboard traversal introduces additional burden. \\
\bottomrule
\end{tabular}
\end{table*}

\paragraph{Component ablation.}
We next asked what information is lost when PSI is reduced to a single proxy. Figure~\ref{fig:component-ablation} compares simplified measures against the PSI component diagnosis across successful deployment traces. Each miss indicates that the proxy fails to capture the dominant burden source identified by the component vector. The result is not that every audit must use scalar PSI, but that the component vector preserves distinctions that single-proxy measures collapse.

\begin{figure}[t]
\centering
\footnotesize
\begin{tikzpicture}
\begin{axis}[
    xbar,
    width=\columnwidth,
    height=5.8cm,
    xmin=0, xmax=55,
    xlabel={Traces with missed burden source (\%)},
    symbolic y coords={
      First-pane visibility,
      Keyboard reachable,
      Time only,
      Hidden reveals only,
      Distance only,
      Clicks only
    },
    ytick=data,
    yticklabel style={font=\sffamily\scriptsize},
    xticklabel style={font=\sffamily\scriptsize},
    xlabel style={font=\sffamily\scriptsize},
    nodes near coords,
    every node near coord/.append style={font=\sffamily\tiny, xshift=2pt},
    axis line style={draw=gray!70},
    tick style={draw=gray!70},
    grid=major,
    grid style={draw=gray!15},
    enlarge y limits=0.13
]
\addplot[
    fill=TableEmph,
    draw=red!55!black
] coordinates {
    (31,First-pane visibility)
    (27,Keyboard reachable)
    (35,Time only)
    (42,Hidden reveals only)
    (39,Distance only)
    (48,Clicks only)
};
\end{axis}
\end{tikzpicture}
\caption{Failure rate of simplified proxy measures relative to PSI component diagnoses across successful deployment traces. A failure means that the proxy misses the dominant burden source identified by the PSI component vector.}
\label{fig:component-ablation}
\end{figure}

Together, these comparisons show that PSI is not merely a replacement for a checklist. Simple heuristics are useful for identifying specific failure modes, but they collapse structurally different burdens. PSI preserves those differences by reporting both a scalar summary and the component vector that produced it.

\subsection{Detector reliability}
Because PSI depends on identifying when a meaningful non-accepting path becomes visible and actionable, we also report detector reliability on two core subtasks: first-viewport visibility and one-click actionability. 

\begin{table}[t]
\centering
\footnotesize
\caption{Human annotation and detector reliability for core PSI subtasks. Highlighting marks strongest reliability and comparatively lower agreement.}
\label{tab:detector-reliability}
\setlength{\tabcolsep}{4.2pt}
\renewcommand{\arraystretch}{1.15}
\begin{tabular}{@{}p{2.65cm}cccc@{}}
\toprule
\rowcolor{ReliabilityHeader}
\textbf{Subtask} & \textbf{$n$} & \textbf{Agreement / $\kappa$} & \textbf{Precision} & \textbf{Recall} \\
\midrule
First-viewport visibility & 100 & 0.88 / 0.80 & \strongcell{0.91} & 0.86 \\
\rowcolor{ReliabilityCaution}
One-click actionability & 100 & 0.84 / \textbf{0.74} & 0.88 & 0.82 \\
Meaningful alternative label & 100 & 0.86 / 0.78 & 0.89 & 0.84 \\
\rowcolor{ReliabilityCaution}
Archetype classification & 100 & 0.82 / 0.76 & 0.87 & \textbf{0.81} \\
Keyboard traversal status & 80 & \strongcell{0.90 / 0.83} & \strongcell{0.93} & \strongcell{0.88} \\
\bottomrule
\end{tabular}
\end{table}

Two auditors independently annotated a stratified sample of consent surfaces drawn from the canonical patterns and live-deployment traces. The annotation covered first-viewport visibility, one-click actionability, meaningful-alternative labels, archetype assignment, and keyboard traversal status. As shown in Table~\ref{tab:detector-reliability}, agreement was substantial across subtasks, with the strongest reliability for keyboard traversal status, the lowest raw agreement for archetype classification, and the lowest $\kappa$ for one-click actionability. The most common disagreements involved euphemistic labels such as ``Manage experience,'' icon-only toggles without accessible names, and controls that appeared visually present but were disabled, occluded, or unreachable under keyboard traversal. These disagreement patterns are consistent with prior work showing that privacy labels, link texts, and privacy-policy language often fail to communicate data practices clearly, especially when users must infer the practical consequence of a label or control \citep{habib2021toggles,rao2016mismatched,balash2024ioslabels,zhang2024expandablegrid}.

We also conducted a small inter-rater reproducibility check for the full PSI audit procedure. Two auditors independently scored 24 consent surfaces sampled across canonical and live-deployment conditions, each identifying the first meaningful alternative, abstracting the event strip, and extracting the component values used for PSI. Agreement was high on the first meaningful alternative ($\kappa=0.82$) and on event-strip abstraction across the five audit event classes ($\kappa=0.79$). The resulting PSI values showed strong consistency across auditors (ICC$=0.91$, two-way random effects, absolute agreement), with disagreements arising primarily in borderline cases involving euphemistic labels or disclosures that appeared visually present but did not yet materially advance refusal or customization.

\paragraph{Rendering robustness.}
We also tested whether canonical PSI separations remain stable under modest rendering perturbations. Across the tested conditions, Co-present Choice remained the lowest-burden pattern, Multi-step Modal remained among the highest-burden patterns, and Scroll Wall and Accordion designs remained intermediate but close because they emphasize different burden sources: traversal distance versus hidden reveals. Absolute PSI values changed modestly, with normalized values ranging from 0.92--1.09 across perturbations. We report the full viewport and animation sensitivity plot in Appendix~\ref{app:rendering-robustness}.

\subsection{Sensitivity to weighting profiles}

Because PSI is a weighted scalar summary, we tested whether the main deployment patterns depend on the equal-weight profile. We generated 1,000 plausible weighting profiles by sampling non-negative weights for \(\alpha,\beta,\gamma,\delta\) from a Dirichlet distribution and rescaling them to sum to four, preserving comparability with the default profile. We also evaluated constrained profiles in which no single component could receive more than half of the total weight. For each profile, we recomputed PSI and measured whether the same qualitative ordering held across archetypes and traversal conditions.

\begin{table}[t]
\centering
\footnotesize
\caption{Random-weight robustness over 1,000 plausible weighting profiles. Values report the percentage of profiles under which each qualitative conclusion holds.}
\label{tab:weight-robustness}
\setlength{\tabcolsep}{4.5pt}
\renewcommand{\arraystretch}{1.15}
\begin{tabular}{@{}p{4.1cm}cc@{}}
\toprule
\textbf{Conclusion tested} & \textbf{Profiles supporting} & \textbf{Interpretation} \\
\midrule
Co-present Choice lowest burden & \cellcolor{RobustStrong}\textbf{98.7\%} & \cellcolor{RobustStrong}Highly stable \\
Multi-step Modal highest burden & \cellcolor{RobustStrong}\textbf{94.1\%} & \cellcolor{RobustStrong}Highly stable \\
Keyboard $>$ pointer traversal & \cellcolor{RobustStrong}\textbf{96.4\%} & \cellcolor{RobustStrong}Highly stable \\
Mobile $>$ desktop layout & \cellcolor{RobustStrong}\textbf{91.8\%} & \cellcolor{RobustStrong}Stable \\
Accordion $>$ Scroll Wall & \cellcolor{RobustMedium}72.6\% & \cellcolor{RobustMedium}Moderately stable \\
Default PSI rank correlation & \textbf{$\rho=0.93$ median} & Strong agreement \\
\bottomrule
\end{tabular}
\end{table}

Table~\ref{tab:weight-robustness} summarizes the robustness check. The main ordering was stable across most profiles: Co-present Choice remained the lowest-burden archetype, Multi-step Modal remained the highest-burden archetype, and keyboard traversal increased median PSI relative to pointer traversal in nearly all sampled profiles. Instability occurred mainly between Scroll Wall and Accordion designs, which is expected because those archetypes emphasize different burden sources: distance versus hidden reveals. We therefore treat scalar PSI as a compact summary, while interpreting close comparisons through the underlying component vectors.

Before summarizing design effects, we use three companion interpretive signals. The Assurance Affordance Index (AAI) refers to visible cues that make the consent process appear complete, safe, or reassuring, such as dominant accept buttons, progress cues, or completion microcopy. The Comprehension Support Index (CSI) refers to design support for understanding and revising choices, such as co-present alternatives, local rationales, and persistent change-consent affordances. We use divergence, \( \mathrm{DIV}=\mathrm{AAI}-\mathrm{CSI} \), as an interpretive signal for cases where reassurance cues increase without corresponding comprehension support. These companion signals are not the primary contribution of this paper, but help interpret whether low or high PSI appears alongside meaningful control.

\subsection{Design-effect matrix}
We summarize directional effects. Arrows indicate the monotone effect on each signal compared to a baseline of otherwise identical alternatives present and available local explanations.

\begin{table}[t]
\centering
\small
\renewcommand{\arraystretch}{1.28} 
\setlength{\tabcolsep}{7pt} 

\caption{Directional effects by structural choice. Shaded groups distinguish burden-raising structures from support-improving structures.}
\label{tab:effects}

\newcommand{\effUp}{\textcolor{teal}{\textbf{$\uparrow$}}}
\newcommand{\effDn}{\textcolor{purple}{\textbf{$\downarrow$}}}
\newcommand{\effNu}{\textcolor{gray}{$\approx$}}

\begin{tabular}{@{} l cccc @{}}
\toprule
\textbf{Structural Choice} & \textbf{PSI} & \textbf{AAI} & \textbf{CSI} & \textbf{DIV} \\
\midrule
\multicolumn{5}{@{}l@{}}{\cellcolor{TableHeader}\textbf{Burden-raising structures}} \\
Alternative offscreen (scroll)          & \effUp & \effNu & \effDn & \effUp \\
Alternative behind accordion            & \effUp & \effNu & \effDn & \effUp \\
Multi-step gating                       & \effUp & \effUp & \effDn & \effUp \\
Keyboard loop before control            & \effUp & \effNu & \effDn & \effUp \\
Dominant Accept (salience $>1.5\times$) & \effNu & \effUp & \effDn & \effUp \\
\midrule
\multicolumn{5}{@{}l@{}}{\cellcolor{TableHeader}\textbf{Support-improving structures}} \\
Local rationale near toggle             & \effNu & \effNu & \effUp & \effDn \\
Persistent ``Change consent''           & \effNu & \effNu & \effUp & \effDn \\
\bottomrule
\end{tabular}
\end{table}

Table~\ref{tab:effects} summarizes the directional role of common structural choices. The purpose of the matrix is not to claim exact universal effect sizes, but to make explicit the monotone design logic implied by the PSI framework: moving alternatives offscreen increases traversal burden; placing them behind disclosure increases hidden-reveal burden; and fragmenting them across steps often increases both time and keyboard fragility. By contrast, local rationales and persistent reversibility increase comprehension support without requiring additional pre-choice burden. This provides designers and auditors with a compact review aid that links concrete interface choices to predictable shifts in burden and support.

\begin{figure}[t]
  \centering
  \includegraphics[width=0.85\linewidth]{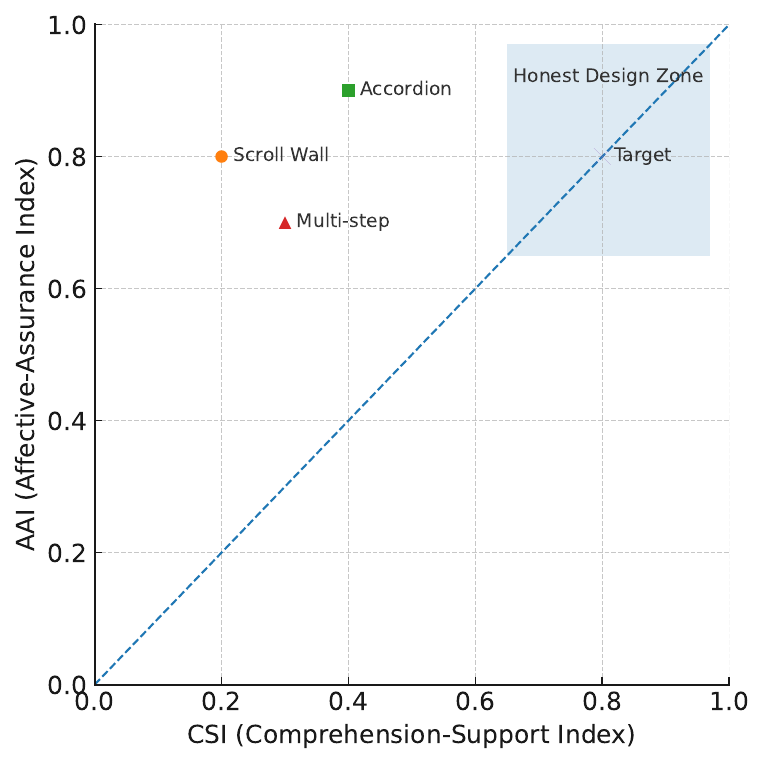}
  \caption{Interpretive divergence map for common consent archetypes. Archetypes that combine stronger reassurance with weaker comprehension support occupy regions of greater divergence, where the interface can appear complete or trustworthy without equally supporting meaningful control.}
  \label{fig:divergence}
\end{figure}

Figure~\ref{fig:divergence} provides an interpretive view of how the canonical archetypes differ in the relation between assurance cues and comprehension support. It should be read together with Table~\ref{tab:effects}: the table summarizes directional design logic, while the figure visualizes how some structures can increase the appearance of completion or trustworthiness without proportionally improving meaningful control.

The figure is not intended to assign exact empirical coordinates to every deployed interface. Rather, it makes visible the structural divergence at the heart of our argument: some archetypes can strengthen reassurance or completion cues while leaving users with weaker support for understanding and exercising alternatives.

\section{Discussion}

\subsection{PSI as a component-first audit measure}
PSI is useful when the design question is not merely whether a consent interface contains a refusal path somewhere, but how much burden is imposed before that path becomes practically available. Read in that way, PSI helps articulate where burden is placed and whether that burden supports meaningful control. The framework is intentionally modest: PSI summarizes pre-choice friction before a meaningful alternative becomes actionable, while companion signals capture visibility, locality, and reversibility. It therefore complements, rather than replaces, studies of comprehension, legal sufficiency, or downstream user welfare.

The deployment results reinforce the value of this component-first interpretation. Interfaces with similar click counts can impose different burdens because the burden may come from viewport displacement, disclosure fragmentation, staged delay, or keyboard fragility. Conversely, two interfaces with similar scalar PSI values may require different redesign responses if their component vectors differ. A high distance term suggests that alternatives should be moved earlier in the flow; a high hidden-reveal term suggests that refusal or customization is fragmented behind disclosure; a high focus-loop term suggests traversal-policy fragility; and a high time term suggests staged or animation-gated delay. In this sense, the scalar PSI score is best understood as a compact summary that points auditors back to the underlying trace rather than as the final interpretation.

This framing also clarifies the role of weighting profiles. The default equal-weight profile provides a transparent first-pass summary, but it should not be read as a claim that one viewport of scrolling, one second of delay, one focus loop, and one hidden reveal have identical psychological effects. Instead, the weights represent declared audit priorities. The random-weight robustness analysis shows whether central conclusions depend on a specific weighting profile, while the component vector preserves the underlying evidence needed for close comparisons.

\subsection{Implications for designers}
For designers, PSI turns a broad complaint about burdensome consent into specific redesign targets. If distance dominates the score, designers can reduce burden by co-presenting refusal or customization options within the first viewport. If hidden reveals dominate, designers can expose category-level controls without requiring nested disclosures. If focus loops dominate, designers can repair tab order, roving \texttt{tabindex}, modal focus management, or keyboard access to settings controls. If time dominates, designers can remove staged animation, debounce delays, or unnecessary intermediate panes. The goal is not to eliminate all interaction from consent flows, but to remove avoidable pre-choice burden before meaningful alternatives become available.

The design-effect matrix and divergence map make this point more concrete. Structural choices such as offscreen alternatives, accordions, and multi-step gating can increase PSI even when legal text remains unchanged. By contrast, local rationales and persistent change-consent affordances improve comprehension support and reversibility without necessarily adding pre-choice burden. This distinction matters because a low-friction design is not automatically a good consent design: meaningful control also requires clear language, local explanations, visible cues, and a route to revisit or revise prior choices, aligning with broader work on how user-facing interface cues shape perceived transparency and control \citep{guo2026feedtaxonomyuserfacingcues}.

To illustrate how PSI can support redesign, Table~\ref{tab:redesign-delta} shows a hypothetical before/after analysis for an anonymized high-burden consent surface. The original interface hides refusal behind a settings pane and introduces a keyboard focus loop after disclosure. The redesign variants progressively co-present the non-accepting alternative, reduce hidden reveals, and add a persistent change-consent affordance.

\begin{table}[t]
\centering
\footnotesize
\caption{Example redesign delta for an anonymized high-burden consent surface. Shading distinguishes the original high-burden interface from lower-burden redesign variants.}
\label{tab:redesign-delta}
\setlength{\tabcolsep}{4.0pt}
\renewcommand{\arraystretch}{1.15}
\begin{tabular}{@{}p{2.05cm}ccccc p{2.7cm}@{}}
\toprule
\rowcolor{TableHeader}
\textbf{Version} & \textbf{$D/\mathrm{vh}$} & \textbf{$T$} & \textbf{$F$} & \textbf{$H$} & \textbf{PSI} & \textbf{Design change} \\
\midrule
Original & 1.4 & 0.9 & 1.0 & 2.0 & \cellcolor{RedesignBefore}\textbf{5.3} & Refusal hidden behind settings; focus loop after reveal \\
\rowcolor{TableStripe}
Redesign A & 0.6 & 0.6 & 0.0 & 1.0 & \textbf{2.2} & Co-present \emph{Reject} with \emph{Accept All}; preserve settings panel \\
Redesign B & 0.4 & 0.5 & 0.0 & 0.0 & \cellcolor{RedesignAfter}\textbf{0.9} & Co-present \emph{Reject} and \emph{Customize}; remove hidden reveal \\
Redesign C & 0.4 & 0.5 & 0.0 & 0.0 & \cellcolor{RedesignBest}\textbf{0.9} & Add local rationale and persistent \emph{Change consent} affordance \\
\bottomrule
\end{tabular}
\end{table}

This example shows why PSI should be interpreted through its components. Redesign A reduces distance and focus burden but leaves one hidden reveal; Redesign B removes the hidden reveal entirely; Redesign C does not further reduce PSI because PSI does not directly measure explanation quality or reversibility. Instead, Redesign C improves companion signals such as comprehension support and reversibility. This distinction prevents PSI from absorbing every desirable property of consent design into one scalar score.

\subsection{Implications for researchers}
For researchers, PSI provides a reproducible structural variable that can be paired with behavioral outcomes in future studies, extending prior work that examines privacy notices, privacy labels, app decisions, and IoT privacy preferences through user-centered evaluation \citep{kelley2010standardizing,kelley2013appdecision,emaminaeini2017iotprivacy,li2022ioslabels,balash2024ioslabels}. Prior work has shown that privacy choices are shaped by interface design, habituation, and comprehension limits; PSI contributes a way to characterize the structure of the consent surface before studying how users respond to it. A controlled study could therefore ask whether higher PSI predicts lower comprehension under fixed exposure time, greater perceived inevitability of acceptance, increased frustration, or lower likelihood of revisiting privacy choices. Such studies would help test the behavioral mechanisms suggested by the privacy-placebo framing without requiring the present audit metric to stand in for user outcomes.

PSI can also support comparative and longitudinal research. Because the metric is derived from first-encounter traces rather than free-form user behavior, researchers can apply the same protocol across device profiles, regions, vendors, or time periods. This makes it possible to study whether consent surfaces become less burdensome after redesigns, policy changes, enforcement actions, or platform-level interventions. In this sense, PSI complements user studies and legal analyses: it does not replace them, but supplies a consistent way to describe the interface structure participants or auditors encounter.

\subsection{Implications for auditors, regulators, and accountability}
For auditors and regulators, PSI can supplement checklist-style compliance review by measuring practical access to alternatives, aligning with policy-facing concerns that dark commercial patterns can undermine nominal choice even when disclosure is formally present \citep{gray2021legalrequirements,luguri2021shining,oecd2022darkcommercial}. A refusal path may exist somewhere in the DOM, behind a settings pane, or in a policy document, but PSI asks when that path becomes visible and actionable under declared traversal conditions. This distinction is important because formal availability does not necessarily imply practical availability. By documenting the distance, delay, focus fragility, and hidden reveals that precede a non-accepting alternative, PSI can help identify interfaces where choice is nominally present but structurally delayed.

PSI may also support accountability work at scale. Auditors could use PSI distributions to compare classes of consent surfaces, identify high-burden outliers for closer review, or track whether redesigns reduce pre-choice burden over time. The metric is especially useful when paired with evidence frames, because the resulting audit record does not only report that an interface scored highly; it shows which structural features produced the score. However, PSI should remain a triage and diagnostic signal rather than a standalone compliance determination. Legal sufficiency, semantic clarity, and user comprehension require additional forms of review.

\subsection{Responsible use and future validation}
A natural next step is a preregistered validation study connecting PSI to broader privacy-choice evaluation frameworks and testing whether lower pre-choice burden corresponds to improved usability, better recall, lower resignation or frustration, and stronger perceived control under controlled conditions \citep{habib2022choiceusability,machuletz2020multiple,pearman2022hipaa,biselli2024personalised,guo2026temporaldriftprivacyrecall}. Such a study should treat PSI as an interface-side predictor rather than as a proxy for comprehension itself.

The broader design principle suggested by this work is simple: alternatives should appear early, explanation should remain local to the relevant control, and reversibility should remain persistently available after the initial choice. Reducing PSI is therefore not about making consent faster at any cost. It is about reducing avoidable pre-choice burden while preserving the conditions needed for meaningful control.

\section{Limitations}
Our lens is structural, which creates several limitations. First, PSI depends on device and rendering conditions: viewport size, zoom, text scaling, and animation timing all influence distance and time terms. For this reason, we report sensitivity to modest viewport and animation changes rather than treating any single absolute score as canonical. Second, PSI depends on a declared traversal policy. Pointer and keyboard paths may diverge substantially, and while this divergence is substantively informative, it also means that reported scores should always be paired with the traversal condition under which they were obtained. Third, identifying a ``meaningful alternative'' depends on labels, roles, visibility, and locality. Euphemistic wording or non-standard controls can therefore mislead heuristic detectors, which is why we pair PSI with annotated first-encounter screenshots and report detector reliability rather than claiming perfect automation. Fourth, consent surfaces change over time and may vary by region, session, or vendor configuration; PSI should therefore be treated as a snapshot unless repeated crawls are used. Finally, PSI diagnoses pre-choice burden, not legal sufficiency, comprehension, or downstream welfare. The metric is best understood as a reproducible audit signal that can support, but not replace, complementary legal and user-centered evaluation.

\subsection{What PSI should not be used for}
PSI should not be used as a standalone measure of legal compliance, user comprehension, user satisfaction, welfare, or organizational privacy ethics. A low PSI score does not prove that a consent surface is legally sufficient or that users understand the consequences of their choices. Likewise, a high PSI score does not by itself prove that a design is unlawful or intentionally deceptive. PSI measures a narrower construct: observable pre-choice burden before a meaningful non-accepting alternative becomes visible and actionable.

PSI should also not be reported as a scalar score without its component vector, weighting profile, traversal policy, and evidence frame. Reporting only a single score can obscure whether burden comes from distance, delay, keyboard fragility, or hidden reveals. This matters because different redesigns are appropriate for different burden sources. Finally, PSI should not be used to reward superficial optimization. A design might reduce scrolling distance while still using euphemistic labels, weak explanations, or poor reversibility. For this reason, PSI should be interpreted alongside semantic review, accessibility checks, comprehension affordances, and legal analysis.

\section{Conclusion}
We introduced \emph{performative scrolling} as a lens for understanding how consent interfaces can produce the appearance of careful review while structuring users toward routine acceptance. To make that structure inspectable, we proposed the \emph{Performative Scrolling Index} (PSI), a reproducible design-side measure of pre-choice burden computed from observable interface signals. Across canonical patterns and a 500-site live deployment, PSI highlighted how offscreen alternatives, hidden disclosure, and keyboard fragility can compound the burden imposed before a meaningful alternative becomes actionable. We do not claim that PSI measures comprehension or legal validity. Rather, its contribution is to make a neglected part of consent design auditable: the friction users must cross before refusal, customization, or revision becomes practically available. By pairing PSI with annotated first-encounter evidence, companion signals, and sensitivity reporting, we hope to support more transparent audits and more honest consent design.

\bibliographystyle{ACM-Reference-Format}
\bibliography{refs}

\appendix
\section{Appendix}

\subsection{Supplementary Discussion}
PSI relates to pre-choice burden, not legal sufficiency or downstream welfare. Accordingly, it may overstate burden when long text is demonstrably optional and honest alternatives are already available on the first pane, while it may also understate burden when refusal is hidden through euphemistic wording or unusually dense presentation. For this reason, we treat PSI as a structural audit signal rather than a standalone normative judgment.

A further implication is that audits should privilege first-encounter states. Repeated exposure can reduce PSI through familiarity alone without yielding any corresponding gain in comprehension support, consistent with findings from warning and permission habituation \citep{felt2012android}. To keep the metric interpretable and resistant to gaming, audits should ideally report not only the final score, but also the shortest annotated path, the underlying event timeline (distance, time, focus loops, hidden reveals), and the weighting profile used. We therefore discourage per-user scoring and instead position PSI as a reproducible interface-side audit measure.

More broadly, PSI should be interpreted together with companion signals. Increases in assurance cues (AAI) should not be taken as beneficial unless they are paired with non-decreasing comprehension support (CSI). This guardrail is consistent with accessibility guidance on focus order and keyboard operability, as well as choice-architecture guidance that alternatives should be co-present and explanations local to the relevant controls \citep{wcag21,johnson2012choicearchitecture}.

\paragraph{Operational rules for AAI and CSI.}
\emph{AAI (Assurance cues):} (i) default accept control salience ratio greater than 1.5$\times$ the nearest meaningful alternative; (ii) progress or celebratory microcopy present (e.g., ``You're all set,'' step counts); and (iii) completion affordance visually dominant within the first pane.

\emph{CSI (Comprehension affordances):} (i) a non-accept alternative is co-present in the first viewport and reachable in one primary interaction; (ii) a local rationale of no more than one sentence appears within 120\,px of the relevant control; and (iii) a persistent ``Change consent'' affordance remains visible from the landing context.

\subsection{Extra Audit Examples}

\begin{table*}[t]
\centering
\footnotesize
\renewcommand{\arraystretch}{1.4} 

\newcommand{\uiaction}[1]{\textcolor{darkgray}{\textsf{#1}}}
\newcommand{\uiflow}{\textcolor{lightgray}{$\rightarrow$}}

\caption{Supplementary audit examples.}
\label{tab:appendix-examples}

\begin{tabular}{@{} 
    >{\raggedright\arraybackslash}p{2.5cm} 
    >{\raggedright\arraybackslash}p{3.8cm} 
    >{\raggedright\arraybackslash}p{2.8cm} 
    >{\raggedright\arraybackslash}p{5.5cm} 
@{}}
\toprule
\textbf{Pattern} & \textbf{Representative Event Strip} & \textbf{Primary PSI Driver} & \textbf{Interpretive Note} \\
\midrule
Accordion-style disclosure & 
\uiaction{expand} \uiflow{} \uiaction{toggle} \uiflow{} \uiaction{action} & 
Hidden reveals ($H$) & 
Burden comes mainly from concealed options rather than from scrolling distance. \\

Multi-step modal & 
\uiaction{action} \uiflow{} \uiaction{action} \uiflow{} \uiaction{toggle} \uiflow{} \uiaction{action} & 
Time + staged gating ($T$, $H$) & 
Burden accumulates across panes before refusal or customization becomes actionable. \\
\bottomrule
\end{tabular}
\end{table*}

To illustrate how PSI distinguishes among structurally different consent surfaces, we provide two additional compact audit examples beyond the worked vignette in the main paper. These examples are not intended as prevalence claims; rather, they show how the same audit procedure yields different burden profiles depending on where and how meaningful alternatives are surfaced.

In an accordion-style disclosure design, the first pane presents a dominant accept control together with a collapsed ``Manage settings'' section. Under the least-effort pointer policy, the first meaningful alternative becomes reachable only after expanding the disclosure and revealing category-level controls. A representative event strip is {\small \texttt{EV\_EXPAND} $\rightarrow$ \texttt{EV\_TOGGLE} $\rightarrow$ \texttt{EV\_ACTION}}. In this case, PSI rises primarily through the hidden-reveal term \(H\), while distance remains relatively low because the relevant controls are technically near the initial viewport. By contrast, in a multi-step modal the first pane provides only an accept option and a continuation affordance leading to a second settings pane. Under the least-effort policy, the first meaningful alternative is not actionable until after at least one staged transition. A representative event strip is {\small \texttt{EV\_ACTION} $\rightarrow$ \texttt{EV\_ACTION} $\rightarrow$ \texttt{EV\_TOGGLE} $\rightarrow$ \texttt{EV\_ACTION}}. Here, PSI increases through both time and hidden structure, and keyboard traversal may further increase burden if focus is reset or trapped across panes. Relative to the accordion example, this pattern illustrates how burden can arise from stepwise gating rather than from disclosure alone.

These supplementary examples reinforce the interpretation developed in the main text: PSI does not merely count clicks or elapsed time, but helps localize why burden increases---whether through concealment, staged interaction, or displacement.

\subsection{Rendering Robustness}
\label{app:rendering-robustness}
We tested whether PSI rankings remain stable under modest rendering changes. Across the canonical patterns, $\pm20\%$ viewport changes and fixed animation delays of $+100/200$\,ms preserved the rank ordering (Scroll Wall $>$ Accordion $>$ Multi-step $>$ Co-present) while changing absolute PSI only modestly. These rankings refer to the canonical pattern instantiations used for sensitivity testing and should not be conflated with the live deployment values reported in Table~\ref{tab:deployment-summary}, where implementation-specific details can change the ordering. This suggests that PSI is best interpreted as a comparative burden measure whose rank ordering is more stable than any single absolute value. For that reason, we recommend reporting the tested viewport and animation profile alongside PSI rather than presenting a score in isolation.

Figure~\ref{fig:viewport-animation-sensitivity} shows that modest rendering changes preserve the relative rank ordering of the canonical patterns.

\begin{figure}[h]
\centering
\footnotesize
\begin{tikzpicture}
\begin{axis}[
    width=\columnwidth,
    height=5.4cm,
    ymin=0.88, ymax=1.12,
    xmin=0.7, xmax=3.3,
    xlabel={Condition},
    ylabel={Relative PSI (baseline = 1.0)},
    xtick={1,2,3},
    xticklabels={Baseline, {$+20\%$ vh}, {$+200$ ms}},
    xticklabel style={font=\sffamily\tiny},
    yticklabel style={font=\sffamily\scriptsize},
    xlabel style={font=\sffamily\scriptsize},
    ylabel style={font=\sffamily\scriptsize},
    legend style={
      font=\sffamily\tiny,
      at={(0.02,0.98)},
      anchor=north west,
      draw=none,
      fill=white
    },
    grid=major,
    grid style={draw=gray!15},
    axis line style={draw=gray!70},
    tick style={draw=gray!70}
]

\addplot[
    mark=*,
    thick,
    color=blue!65!black
] coordinates {
    (1,1.00) (2,0.92) (3,1.05)
};
\addlegendentry{Scroll Wall}

\addplot[
    mark=square*,
    thick,
    color=orange!80!black
] coordinates {
    (1,1.00) (2,0.93) (3,1.06)
};
\addlegendentry{Accordion}

\addplot[
    mark=triangle*,
    thick,
    color=purple!70!black
] coordinates {
    (1,1.00) (2,0.96) (3,1.08)
};
\addlegendentry{Multi-step}

\addplot[
    mark=diamond*,
    thick,
    color=green!50!black
] coordinates {
    (1,1.00) (2,0.94) (3,1.09)
};
\addlegendentry{Co-present}

\end{axis}
\end{tikzpicture}
\caption{Sensitivity of PSI to viewport and animation perturbations, normalized to each pattern's baseline condition. Increasing viewport height slightly reduces PSI across patterns, while adding a 200\,ms delay slightly increases PSI. Qualitative ordering remains unchanged.}
\label{fig:viewport-animation-sensitivity}
\end{figure}

\subsection{Additional Edge Cases}
The main text defines the decision rules used for ambiguous labels, route selection, focus behavior, and effective viewport calculation. Here we provide additional edge-case examples to clarify how the protocol should be applied in borderline situations. These examples are intended as audit decision cases rather than prevalence claims.

\begin{table*}[t]
\centering
\footnotesize
\caption{Additional edge-case decision examples for applying the PSI audit protocol. These examples clarify how ambiguous controls, route selection, focus behavior, and viewport calculation should be handled.}
\label{tab:appendix-edge-cases}
\setlength{\tabcolsep}{4.5pt}
\renewcommand{\arraystretch}{1.18}
\begin{tabular}{@{}p{2.9cm}p{4.2cm}p{3.2cm}p{4.7cm}@{}}
\toprule
\rowcolor{TableHeader}
\textbf{Edge case} & \textbf{Observed interface pattern} & \textbf{Audit decision} & \textbf{Rationale} \\
\midrule

Ambiguous ``Learn more'' control &
The first pane shows \emph{Accept All} and \emph{Learn more}. Activating \emph{Learn more} opens explanatory text but does not reveal refusal, customization, or saving controls. &
Do not count as a meaningful alternative. &
The control provides information but does not materially advance the user toward refusing, narrowing, or revising consent. \\

\rowcolor{TableStripe}
``Manage experience'' reveals settings &
The first pane shows \emph{Accept All} and \emph{Manage experience}. Activating it opens a settings panel with category toggles and a \emph{Save choices} button. &
Count as a meaningful alternative only after the settings panel becomes visible and actionable. &
The label is euphemistic, but the activated route does advance toward narrowing consent. The reveal contributes to $H$. \\

Partially visible button &
A \emph{Reject All} button is partially visible at the bottom edge of a scrollable consent surface, but its label is clipped and the button cannot be activated until the surface is scrolled. &
Do not count as visible/actionable at first encounter; include the required scroll distance. &
A DOM-present or partially rendered control is insufficient if the user must scroll before the alternative is legible and actionable. \\

\rowcolor{TableStripe}
Disabled save button &
A settings panel contains a \emph{Save choices} button, but the button is disabled until at least one toggle is changed. &
Do not count the disabled button as actionable; include the toggle action needed to enable it. &
Actionability requires that the control can be activated immediately under the declared traversal policy. \\

Policy-link detour &
The banner contains \emph{Privacy Policy} beside \emph{Accept All}. The policy page describes opt-out rights but does not provide a direct control in the consent surface. &
Do not count the policy link as a meaningful alternative. &
The audit target is practical access to a non-accepting action, not access to general explanatory or legal text. \\

\rowcolor{TableStripe}
Icon-only settings button &
A gear icon appears beside \emph{Accept All}. It has no visible label, but an accessible name of ``Cookie settings'' and opens refusal controls. &
Count as a route to a meaningful alternative if reachable under the traversal policy; record the reveal. &
The control is actionable and semantically identifiable through its accessible name, but the hidden settings panel still contributes to $H$. \\

Focus trap before settings &
Keyboard traversal cycles between \emph{Accept All} and close controls without reaching \emph{Manage settings}, while pointer traversal can open settings. &
Count a focus-loop event under keyboard traversal and flag traversal-policy fragility. &
The pointer path is not sufficient evidence of practical access under keyboard traversal. \\

\rowcolor{TableStripe}
Legitimate disclosure focus shift &
Activating \emph{Customize} moves keyboard focus into the newly opened settings panel and lands on the first category toggle. &
Do not count as a focus loop. &
The focus movement advances the route toward a visible and actionable non-accepting control. \\

Nested consent-surface scroll &
The page itself does not scroll, but the consent modal has an internal scroll container. The refusal control appears below the modal fold. &
Compute $D/\mathrm{vh}$ relative to the effective consent-surface viewport. &
The burden occurs inside the consent surface, so the relevant viewport is the scrollable modal region rather than the full browser window. \\

\rowcolor{TableStripe}
Multiple visible non-accepting paths &
Both \emph{Reject All} and \emph{Customize} are visible on the first pane. \emph{Reject All} is immediately actionable, while \emph{Customize} opens a longer settings flow. &
Follow the shortest direct non-accepting path, usually \emph{Reject All}. &
The least-effort protocol measures burden before the first meaningful non-accepting alternative becomes actionable. \\

\bottomrule
\end{tabular}
\end{table*}

These examples clarify that PSI is not triggered merely by the presence of additional text, links, or DOM elements. A control counts only when it provides practical access to refusing, narrowing, or revising consent under the declared traversal policy. Conversely, the protocol does not penalize every intermediate interaction: disclosure, focus relocation, or scrolling is counted only when it delays the first visible and actionable non-accepting alternative.

\begin{figure}[t]
  \centering
  \includegraphics[width=0.85\linewidth]{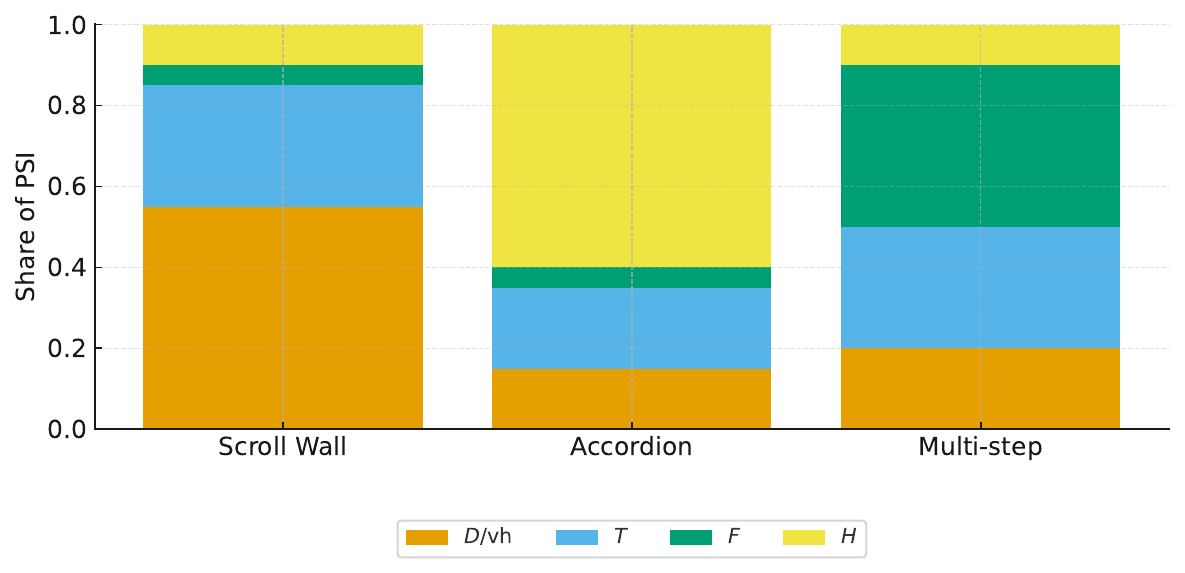}
  \caption{Normalized PSI component shares by archetype. Scroll Wall emphasizes distance and time; Accordion emphasizes hidden reveals; Multi-step introduces focus fragility.}
  \label{fig:appendix-psi-components-v2}
\end{figure}

\subsection{Future Work}
\begin{figure}[t]
  \centering
  \includegraphics[width=0.85\linewidth]{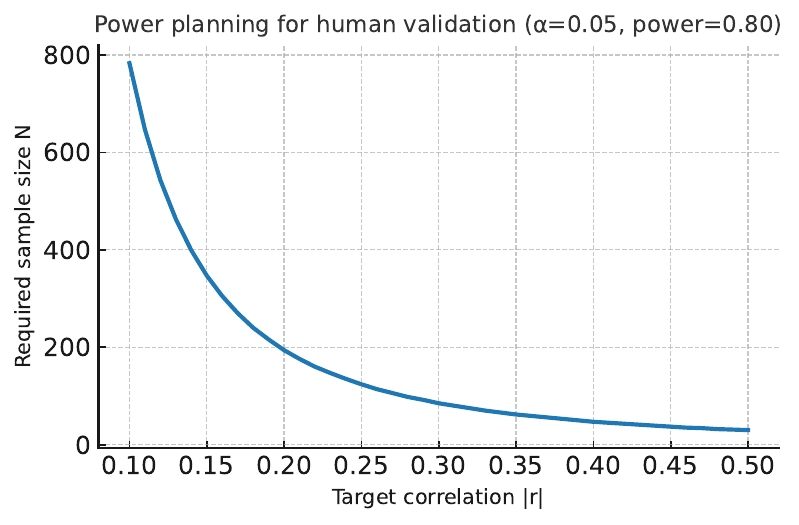}
  \caption{Sample size required to detect a correlation under a two-tailed test with $\alpha{=}0.05$ and power $=0.80$.}
  \label{fig:appendix-power-v2}
\end{figure}

A natural next step is a preregistered validation study testing whether PSI predicts (H1) lower comprehension under a fixed exposure interval, (H2) higher resignation or frustration, such as a perceived inevitability of acceptance, and (H3) whether co-present alternatives reduce PSI and divergence without reducing completion. Power planning appears tractable: detecting $|r|{=}0.25$ with $\alpha{=}0.05$ and 80\% power requires approximately 120 participants, while detecting $|r|{=}0.30$ requires approximately 85 participants (Figure~\ref{fig:appendix-power-v2}).

At a larger scale, PSI could support deployed audits across device profiles and regions. A headless crawler could execute the least-effort policy, extract PSI/AAI/CSI signals from DOM and timing traces, and report distributional variation across vendors or categories. A second-pass fairness or accessibility audit could then re-run the same surfaces under keyboard-only traversal or under stricter locality thresholds. These extensions would help connect interface-side burden measurement to broader monitoring of privacy-choice environments.

The appendix component view in Figure~\ref{fig:appendix-psi-components-v2} also suggests a practical redesign workflow. Interventions such as co-present Reject/Customize controls, one-sentence local rationales, and a persistent Change consent affordance should reduce PSI while increasing CSI. Visualizing before/after deltas at the component level may therefore help designers and auditors identify which structural changes most directly improve meaningful control.

\paragraph{Scaling.}
A next step is a larger crawl (e.g., top-$K$ sites within a category) to characterize PSI distributions across vendors and regions, quantify detector failure modes at scale, and examine how burden varies across device conditions and traversal policies.

\paragraph{Toward robust semantic detection.}
Our current detector is rule-based and remains vulnerable to euphemism and non-standard labeling. One practical extension is a two-stage pipeline: deterministic role/label heuristics first, followed by an LLM-based classifier only for uncertain cases, constrained to a fixed label set such as ACCEPT / REJECT / SETTINGS / UNKNOWN. Such a pipeline could improve recall on euphemistic controls while preserving interpretability and enabling precision/recall auditing on held-out screenshots.

\end{document}